\documentclass[11pt,titlepage]{article}
\usepackage{latexsym}
\usepackage{amsfonts}
\usepackage{amssymb}

\usepackage{times}
\usepackage{amsmath}

\mathcode`\0="0030      %
\mathcode`\1="0031
\mathcode`\2="0032
\mathcode`\3="0033
\mathcode`\4="0034
\mathcode`\5="0035
\mathcode`\6="0036
\mathcode`\7="0037
\mathcode`\8="0038
\mathcode`\9="0039

\makeatletter
\def\@begintheorem#1#2{\trivlist\item[\hskip\labelsep{\bf #1\ #2}]}
\def\foobarpt{\textfont\z@\tenrm 
  \scriptfont\z@\ninrm \scriptscriptfont\z@\sevrm
\textfont\@ne\tenmi \scriptfont\@ne\ninmi \scriptscriptfont\@ne\sevmi
\textfont\tw@\tensy \scriptfont\tw@\ninsy \scriptscriptfont\tw@\sevsy
\textfont\thr@@\tenex \scriptfont\thr@@\tenex \scriptscriptfont\thr@@\tenex
\def\unboldmath{\everymath{}\everydisplay{}\@nomath\unboldmath
          \textfont\@ne\tenmi 
          \textfont\tw@\tensy \textfont\lyfam\tenly
          \@boldfalse}\@boldfalse
\def\boldmath{\@ifundefined{tenmib}{\global\font\tenmib\@mbi\@magscale1\global
        \font\tensyb\@mbsy \@magscale1\global\font
         \tenlyb\@lasyb\@magscale1\relax\@addfontinfo\@xiipt
              {\def\boldmath{\everymath
                {\mit}\everydisplay{\mit}\@prtct\@nomathbold
                \textfont\@ne\tenmib \textfont\tw@\tensyb 
                \textfont\lyfam\tenlyb\@prtct\@boldtrue}}}{}\@xiipt\boldmath}%
\def\prm{\fam\z@\tenrm}%
\def\pit{\fam\itfam\tenit}\textfont\itfam\tenit \scriptfont\itfam\ninit
   \scriptscriptfont\itfam\sevit
\def\psl{\fam\slfam\tensl}\textfont\slfam\tensl 
     \scriptfont\slfam\tensl \scriptscriptfont\slfam\tensl
\def\pbf{\fam\bffam\tenbf}\textfont\bffam\tenbf 
   \scriptfont\bffam\ninbf \scriptscriptfont\bffam\ninbf 
\def\ptt{\fam\ttfam\tentt}\textfont\ttfam\tentt
   \scriptfont\ttfam\nintt \scriptscriptfont\ttfam\nintt 
\def\psf{\fam\sffam\tensf}\textfont\sffam\tensf
    \scriptfont\sffam\tensf \scriptscriptfont\sffam\tensf
\def\psc{\@getfont\psc\scfam\@xiipt{\@mcsc\@magscale1}}%
\def\ly{\fam\lyfam\tenly}\textfont\lyfam\tenly 
   \scriptfont\lyfam\ninly \scriptscriptfont\lyfam\sevly
 \@setstrut \rm}

\newcommand{\singlespacing}{\let\CS=
\@currsize\renewcommand{\baselinestretch}{1}\tiny\CS}
\newcommand{\singlespacingplus}{\let\CS=
\@currsize\renewcommand{\baselinestretch}{1.25}\tiny\CS}
\newcommand{\doublespacing}{\let\CS=
\@currsize\renewcommand{\baselinestretch}{1.75}\tiny\CS}
\newcommand{\draftspacing}{\let\CS=
\@currsize\renewcommand{\baselinestretch}{2.0}\tiny\CS}

\newcommand{\niceonespacing}{\let\CS=\@currsize\renewcommand{\baselinestretch}{1.1}\tiny\CS}
\newcommand{\nicetwospacing}{\let\CS=\@currsize\renewcommand{\baselinestretch}{1.2}\tiny\CS}
\newcommand{\nicefoospacing}{\let\CS=\@currsize\renewcommand{\baselinestretch}{1.18}\tiny\CS}
\newcommand{\nicethreespacing}{\let\CS=\@currsize\renewcommand{\baselinestretch}{1.3}\tiny\CS}
\newcommand{\singlespacingplusplus}{\let\CS=\@currsize\renewcommand{\baselinestretch}{1.35}\tiny\CS}
\newcommand{\nicefivespacing}{\let\CS=\@currsize\renewcommand{\baselinestretch}{1.5}\tiny\CS}
\newcommand{\nicesixpacing}{\let\CS=\@currsize\renewcommand{\baselinestretch}{1.6}\tiny\CS}

\makeatother

\makeatletter
\def\@cite#1#2{[#1\if@tempswa , #2\fi]}
\makeatother

\newcommand\seq{\subseteq}
\renewcommand\.{\cdot}
\newcommand\<{\langle}
\renewcommand\>{\rangle}

\newcommand{\sigmastar}{\mbox{$\Sigma^\ast$}}
\newcommand{\equalsdef}{\stackrel{\mbox{\protect\scriptsize df}}{=}}

\newcommand\N{{\rm I\!N}}

\newcommand\p{\mbox{\rm P}}

\newcommand\fp{\mbox{\rm FP}}
\newcommand\np{\mbox{\rm NP}}

\newcommand\conp{\mbox{\rm coNP}}
\newcommand\coup{\mbox{\rm coUP}}

\newcommand\fewp{\mbox{\rm FewP}}
\newcommand\cofewp{\mbox{\rm coFewP}}

\newcommand\up{{\mbox{UP}}}

\newcommand\easyforallio{\mbox{EASY}^{\forall}_{\mbox{\protect\scriptsize io}}}
\newcommand\easyforall{\mbox{EASY}^{\forall}_{\forall}}

\newcommand\easyexistsio{\mbox{EASY}^{\exists}_{\mbox{\protect\scriptsize io}}}
\newcommand\easyexists{\mbox{EASY}^{\exists}_{\forall}}

\newtheorem{theorem}{Theorem}[section]
\newtheorem{corollary}[theorem]{Corollary}

\newtheorem{proposition}[theorem]{Proposition}
\newtheorem{definition}[theorem]{Definition}

\newenvironment{block}{\begin{list}{\hbox{}}{\leftmargin 1em
    \itemindent -1em \topsep 0pt \itemsep 0pt \partopsep 0pt}}{\end{list}}

\begin{document}

\bibliographystyle{alpha}

\title{Characterizations of the Existence of Partial and Total One-Way
Permutations\thanks{An extended abstract 
of this paper was presented at 
the Third Italian Conference on Algorithms and
Complexity~\protect\cite{hem-rot-wec:c:hard-certificates-plus-permutations}.}}

\author{ 
{\em  J\"org Rothe\/}\thanks{
Supported in part 
by grant
NSF-INT-9513368/\protect\linebreak[0]DAAD-315-PRO-fo-ab and
by a NATO Postdoctoral Science Fellowship
from the Deut\-scher Aka\-de\-mi\-scher Aus\-tausch\-dienst
(``Ge\-mein\-sames Hoch\-schul\-sonder\-pro\-gramm~III
von Bund und L\"andern'').
Current address: Department of Computer Science, 
University of Rochester, 
Rochester, NY 14627, USA.
Work done in part while visiting the University of Rochester and 
Le~Moyne College.
} \\
Institut f\"ur Informatik \\
Friedrich-Schiller-Universit\"at Jena \\
07743 Jena, Germany \\
rothe@informatik.uni-jena.de
\and
{\em  Lane A. Hemaspaandra\/}\thanks{Supported 
in part by grants
NSF-INT-9513368/DAAD-315-PRO-fo-ab
and 
NSF-CCR-9322513.
Work done in part while 
visiting 
Friedrich-Schiller-Universit\"at Jena.} 
\\
Department of Computer Science \\
University of Rochester \\
Rochester, NY 14627 \\
lane@cs.rochester.edu
}

\newcount\hour  \newcount\minutes  \hour=\time  \divide\hour by 60
\minutes=\hour  \multiply\minutes by -60  \advance\minutes by \time
\def\mmmddyyyy{\ifcase\month\or Jan\or Feb\or Mar\or Apr\or May\or Jun\or Jul\or
  Aug\or Sep\or Oct\or Nov\or Dec\fi \space\number\day, \number\year}
\def\hhmm{\ifnum\hour<10 0\fi\number\hour :%
  \ifnum\minutes<10 0\fi\number\minutes}
\def\Draft{{\it Draft of \mmmddyyyy}}

\date{}

\typeout{WARNING:  BADNESS used to suppress reporting!  Beware!!}
\hbadness=3000%
\vbadness=10000 %

\setcounter{footnote}{0}
{\singlespacing\maketitle}

\begin{center}
{\large\bf Abstract}
\end{center}
\begin{quotation}
{\singlespacing

\noindent 
In this note, we study the easy certificate classes introduced by
Hemaspaandra, Rothe, and
Wechsung~\cite{hem-rot-wec:jtoappear:hard-certificates}, with regard to the
question of whether or not surjective one-way functions exist. This is
an important open question in cryptology.  We show that the existence
of partial one-way permutations can be characterized by separating P
from the class of UP sets that, for all unambiguous polynomial-time
Turing machines accepting them, always have easy (i.e.,
polynomial-time computable) certificates. This extends results of 
  Grollmann and Selman~\cite{gro-sel:j:complexity-measures}.
  By Gr\"adel's recent results about one-way
  functions~\cite{gra:j:one-way}, this also links statements about
  easy certificates of NP sets with statements in finite model theory.
  Similarly, there exist surjective poly-one one-way functions if and
  only if there is a set $L$ in P such that not all FewP machines
  accepting $L$ always have easy certificates.
  We also establish a condition necessary and sufficient for the
  existence of (total) one-way permutations.

 }
\end{quotation}

\setcounter{page}{1}
\pagestyle{plain}

\sloppy

\section{Introduction}

What makes NP-complete problems intractable?  One possible 
source of their potential intractability is the fact
that there are many possible sets of solutions:
The search space is exponential so the cardinality
of the set of sets of 
solutions is double-exponential in the input size.
Another possible source of NP's complexity is that all solutions
(even if there are just a few of them) may be random in the sense of
Kolmogorov complexity and thus hard to find. For both reasons one
may try to ``remove'' the difficulty from NP by considering 
subclasses of NP that, by definition, contain only easy sets with
respect to either type of difficulty. 
NP's subclasses UP (unambiguous polynomial 
time)~\cite{val:j:checking}
and FewP
(ambiguity-bounded polynomial 
time)~\cite{all:coutdatedExceptForPUNCstuff:complexity-sparse,all-rub:j:print} 
both implicitly reduce the richness of the class of potential solutions
to $2^{n^{{\cal O}(1)}}$.  
To
single out those NP sets that, for all NP machines accepting them,
have easy solutions---i.e., solutions of small Kolmogorov
complexity---for all instances in the set, Hemaspaandra, 
Rothe, and Wechsung~\cite{hem-rot-wec:jtoappear:hard-certificates} 
defined the class $\easyforall$ (see the
next section for precise definitions).  Interestingly, both these
concepts of easy NP sets (to wit, UP and $\easyforall$) have their own
connection to the invertibility of certain types of one-way functions,
as will be stated below.  Intuitively, a one-way function is a
function that is easy to compute but hard to invert. One-way functions
play a central role in
complexity-theoretic
cryptography~\cite{gro-sel:j:complexity-measures}, where the open
question of whether such functions do or do not exist is of central
importance.

It is well-known that many-one one-way functions exist if and only if
$\p \neq \np$. Thus, we cannot hope for an ultimate solution to
the question of whether or not one-way functions exist unless we can
solve the famous $\p \stackrel{\mbox{?}}{=} \np$ question. All we can
hope for is to characterize the existence of certain special types of
one-way functions via complexity-theoretic statements such as the
collapse or separation of the corresponding complexity classes.  Many
types of one-way functions have been studied in the literature.  Most
notable among such results is Grollmann and Selman's characterization
of the existence of certain types of {\em injective\/} one-way
functions by conditions such as $\p \neq \up$ or $\p \neq \up \cap
\coup$~\cite{gro-sel:j:complexity-measures} (see
also~\cite{ko:j:operators}).  Allender 
extended their results
by proving that {\em poly-one\/} one-way functions exist if and only
if \mbox{$\p \neq 
\fewp$}~\cite{all:coutdatedExceptForPUNCstuff:complexity-sparse}.
Watanabe showed that {\em constant-one\/} one-way functions exist if
and only if injective one-way functions
exist~\cite{wat:j:hardness-one-way}, 
notwithstanding the fact that even at the 
level of constant injectivity
it has been shown~\cite{hem-hem:j:quasi} 
that greater injectivity yields strictly more general
reductions.
Watanabe also showed that the existence of
{\em randomized injective\/} one-way functions and the existence of
{\em extensible injective\/} one-way functions, respectively, can be
characterized by the separations $\mbox{BPP} \neq
\mbox{UP}^{\mbox{\scriptsize BPP}}$ and ${\cal P} \neq {\cal
  UP}$~\cite{wat:j:1tt}, where BPP denotes bounded
probabilistic polynomial time~\cite{gil:j:probabilistic-tms},
${\cal P}$ is the class of polynomial-time solvable promise problems
(in the sense
of~\cite{eve-yac:c:promise,eve-sel-yac:j:promise-problems,gro-sel:j:complexity-measures},
see also~\cite{hem-rot:j:boolean}), and ${\cal UP}$ is the class
of unambiguous promise problems.  Finally, Fenner et 
al.~\cite{fen-for-nai-rog:c:inverse}
proved the
existence of {\em surjective many-one\/} one-way functions equivalent
to \mbox{$\p \not\seq \easyforall$}.

In this note, a  characterization of the existence of {\em
  injective and surjective\/} one-way functions is given by separating
P from a class, denoted $\easyforall (\up)$, which combines the
restriction of unambiguous computation with the constraint required by
$\easyforall$.  Thus, $\easyforall (\up)$ {\em simultaneously\/}
reduces the solution space of NP problems to at most one solution and
requires that this one solution can be found and printed out in
polynomial time, if it exists. Furthermore, the existence of {\em
  surjective poly-one\/} one-way functions is shown to be equivalent
to the separation of P and $\easyforall (\fewp)$ (which is the
polynomially ambiguity-bounded analog of $\easyforall (\up)$).
Our work is connected to the seemingly 
unrelated field 
of (finite model) logic; from
Gr\"adel's~\cite{gra:j:one-way} recent results about one-way
functions,
we obtain as a corollary equivalences
between statements about easy certificates of NP sets and statements
in finite model theory such as that the weak definability principle
in a logic on finite structures fails to hold.  In addition, based on
Selman's analogous observation for $\easyforall$ 
(as cited in~\cite{hem-rot-wec:jtoappear:hard-certificates}), 
further characterizations of
the existence of surjective one-way functions are provided in terms of
the question of whether $\easyforall (\up)$ and $\easyforall (\fewp)$
are closed under complementation. Finally, we show that the existence
of total injective one-way functions with a P-rankable
range is a condition necessary and sufficient for the existence of
one-way permutations.

\section{Preliminaries}

All sets considered are subsets of $\sigmastar$, where $\Sigma =
\{0,1\}$.  Functions map from $\sigmastar$ to $\sigmastar$ and are
many-one and partial (unless explicitly specified to be one-one or
total). The length of a string $x \in \sigmastar$ is denoted by $|x|$
and the cardinality of a set $L \seq \sigmastar$ by $\|L\|$. Let
$\epsilon$ denote the empty string. Let $\< \. , \.\>$ be a standard
easily computable pairing 
function (i.e., a bijection between $\sigmastar \times \sigmastar$ and
$\sigmastar$) that can be extended to encode tuples of strings by one
string as usual. Let $\leq_{\mbox{\protect\scriptsize lex}}$ denote
the standard quasi-lexicographical ordering on~$\sigmastar$. For 
each
(single-valued, partial or total) 
function~$f:\sigmastar \rightarrow \sigmastar$, let $\mbox{dom}(f)$
and $\mbox{range}(f)$ denote the domain and range of~$f$,
respectively.

Let NPM be a shorthand for ``nondeterministic polynomial-time Turing
machine.'' For each NPM~$M$, $L(M)$ denotes the language accepted
by~$M$. For each NPM~$M$ and any input~$x$, we denote the set of
accepting paths of $M(x)$ by $\mbox{acc}_{M}(x)$.  An NPM $M$ is a
UP machine (FewP machine, respectively) if, for all inputs $x$, $M(x)$
has at most one (at most polynomially in~$|x|$, respectively)
accepting paths, and $M$ accepts $x$ if and only if $M(x)$ has at least
one accepting path.  UP~\cite{val:j:checking} (respectively,
FewP~\cite{all:coutdatedExceptForPUNCstuff:complexity-sparse,all-rub:j:print}) 
is the class of sets $L$
such that $L = L(M)$ for some UP machine (FewP machine)~$M$. FP
denotes the class of polynomial-time computable functions.
$\easyforall$~\cite{hem-rot-wec:jtoappear:hard-certificates} is defined to be the
class of all sets $L$ for which all NPMs accepting $L$ always (i.e.,
on all inputs $x \in L$) have easy certificates (i.e., accepting paths
whose encoding can be printed in polynomial time). Of the four classes
$\easyforall$, $\easyforallio$, $\easyexists$, and $\easyexistsio$
considered by Hemaspaandra, Rothe,
and Wechsung~\cite{hem-rot-wec:jtoappear:hard-certificates}, only $\easyforall$ is
relevant for the characterization of one-way functions. 

Now let us formally define the UP and FewP analog of $\easyforall$.
Though it is clear that a more general definition of the form
$\easyforall({\cal C},\, {\cal F})$ for complexity classes ${\cal C}$
other than NP, UP, or FewP and for function classes ${\cal F}$ other
than FP can analogously be obtained, we will only define the classes
of interest here.

\begin{definition}\label{def:easy-up}
  For ${\cal C} \in \{ \mbox{NP}, \mbox{UP}, \mbox{FewP} \}$, define
  $\easyforall ({\cal C})$ to be the class of all sets $L$ that either
  are finite, or that satisfy
  (a)~$L \in {\cal C}$, and~(b) for every ${\cal
    C}$-machine $N$ such that $L(N) = L$, there exists an FP function
  $f_N$ such that, for all $x \in L$, $f_{N}(x) \in
  \mbox{acc}_{N}(x)$.
\end{definition}

The inclusions summarized in Proposition~\ref{prop:incl-easy} below
follow immediately from the definition. For instance, the inclusion
\mbox{$\easyforall \seq \easyforall (\fewp)$} holds, since each
$\easyforall$ set $L$ is in P
(see~\cite[Figure~1]{hem-rot-wec:jtoappear:hard-certificates}) and 
thus in FewP, and
moreover since if every NPM accepting $L$ always has easy
certificates, then so does every FewP machine. The inclusion
\mbox{$\easyforall (\up) \seq \p$} holds due to \mbox{$\easyforall
  (\up) \seq \easyexists (\up) = \easyexists = \p$}
(see~\cite[Theorem~2.2.1]{hem-rot-wec:jtoappear:hard-certificates}), 
where $\easyexists
(\up)$ denotes the analog of $\easyforall (\up)$ such that condition
(b) in Definition~\ref{def:easy-up} above is required to hold only for
{\em some\/} UP machine~$N$.

\begin{proposition}\label{prop:incl-easy}
  $\easyforall \seq \easyforall (\fewp) \seq \easyforall (\up) \seq \p
  \seq \up \seq \fewp \seq \np$.
\end{proposition}

Next, we define the types of one-way functions considered in this
paper.  Note that the honesty of one-way functions is required in
order to avoid the case that the FP-noninvertibility is trivial.

\begin{definition}
\begin{enumerate}
\item A function $f$ is {\em honest\/} if there is a polynomial $p$
  such that for every $y \in \mbox{range}(f)$ and for every $x \in
  \mbox{dom}(f)$, if $y = f(x)$ then $|x| \leq p(|y|)$.

\item A function $f$ is {\em poly-one\/} if there is a polynomial $p$
  such that $\|f^{-1}(y)\| \leq p(|y|)$ for each $y \in
  \mbox{range}(f)$.
  
\item A (many-one) function $f$ is said to be {\em $\fp$-invertible\/}
  if there is a function $g \in \fp$ such that for every $y \in
  \mbox{range}(f)$, $g(y)$ prints {\em some\/} value of $f^{-1}(y)$.
  In particular, if $f$ is one-one, $\fp$-invertibility of $f$ means
  $f^{-1} \in \fp$. 

\item A function $f$ is said to be a {\em one-one\/} 
  (respectively, {\em
    poly-one}, {\em many-one\/}) {\em one-way function\/} if $f$ is
  honest, one-one (respectively, 
  poly-one, many-one), $f \in \fp$, and $f$ is not
  FP-invertible. If $f : \sigmastar \rightarrow \sigmastar$ is a
  total, surjective, and one-one one-way function, $f$ is called a
  {\em one-way permutation}. 
\end{enumerate}
\end{definition}

Sometimes the following weaker definition of honesty is used: $f$ is
{\em honest\/} if there is a polynomial $p$ such that for every $y \in
\mbox{range}(f)$ there is a string $x \in \mbox{dom}(f)$ such that 
$y = f(x)$ and $|x| \leq p(|y|)$.   
All claims in this paper, except those 
involving weak one-way functions (defined later), hold also for 
this alternate definition.

Note that we discuss one-way functions in the complexity-theoretic
setting introduced by Grollmann and
Selman~\cite{gro-sel:j:complexity-measures}. So-called cryptographic
one-way functions are not discussed here, though we should mention
that one-way permutations have been interestingly studied in that
context~\cite{yao:c:trapdoor,imp-rud:c:one-way-perms,has-imp-lev-lub:t:psgen-oneway}.

\section{Characterizing the Existence of Surjective 
One-Way Functions}\label{s:sur-one}

Fenner et al.~\cite{fen-for-nai-rog:c:inverse} have
characterized the existence of surjective {\em many-one\/} one-way
functions by the condition $\p \not\seq \easyforall$. 
In this section, we give analogous characterizations of the existence
of surjective one-one one-way functions and surjective poly-one
one-way functions by separating P from $\easyforall (\up)$ and
$\easyforall (\fewp)$, respectively.
We mention here that
there is a relativization 
in which $\p \not\seq \easyforall$
does not imply $\p \neq \np \cap
\conp$~(\cite{nao-imp:c:decision-trees},
see also
\cite{cre-sil:c:sperner,for-rog:c:separability,fen-for-nai-rog:c:inverse}).

We begin with the characterization of the existence of surjective
one-one one-way functions.  
Note that the type of 
function discussed in item~(2) of Theorem~\ref{thm:up} below 
is the partial-function analog of a 
(total) one-way permutation.  
Note also that the equivalence of statements~(1),~(3), 
and~(4) in Theorem~\ref{thm:up} holds in analogy to the
case of $\easyforall$
(see~\cite{hem-rot-wec:jtoappear:hard-certificates,fen-for-nai-rog:c:inverse}).

\begin{theorem}
\label{thm:up}
The following are equivalent.
\begin{enumerate}
\item $\easyforall (\up) \neq \p$.

\item There exists a partial one-one one-way function $f$ with
  $\mbox{range}(f) = \sigmastar$.

\item $\sigmastar \not\in \easyforall (\up)$.

\item \label{selman-item} $\easyforall (\up)$ 
is not closed under complementation.
\end{enumerate}
\end{theorem}

\noindent {\bf Proof.} \quad 
Clearly, (3) implies (4), since $\overline{\sigmastar} = \emptyset$ as
a finite set is in $\easyforall (\up)$. (4) 
immediately implies (1). To
see that (1) implies (3), assume there is a set $L \in \p$ such that
$L \not\in \easyforall (\up)$. Let $N$ be some UP machine accepting
$L$ such that no FP function exists that outputs the accepting path of
$N(x)$ for all inputs~$x \in L$. Let $M$ be some P machine that
accepts~$\overline{L}$.  Consider the following NPM~$N'$: On
input~$x$, $N'$ guesses whether $x \in L$ or $x \in \overline{L}$. If
the guess was ``$x \in L$,'' $N'$ simulates $N(x)$; otherwise, it
simulates~$M(x)$. Then, $N'$ is a UP machine accepting~$\sigmastar$.
Note that the accepting computation of $N'(x)$ for inputs $x \in L$
contains the accepting computation of~$N(x)$. Since $L$ cannot be empty
(in fact, $L$ cannot be finite, for otherwise we would
have had $L \in
\easyforall (\up)$), no FP function can output, for all inputs $x \in
\sigmastar$, the accepting path of~$N'(x)$. Hence, $\sigmastar \not\in
\easyforall (\up)$.

\smallskip

(3) implies (2): Assume $\sigmastar \not\in \easyforall (\up)$. Let
$M$ be a UP machine accepting $\sigmastar$ such that no FP function
can output the accepting path of $M(y)$ for all~$y \in \sigmastar$. For
any input~$y$, let $\mbox{comp}_{M}(y)$ denote the unique accepting
path (encoded as a sequence of configurations) of~$M(y)$.  As
in~\cite{gro-sel:j:complexity-measures}, define the function $f$ to be
\[
f(x) \equalsdef 
\left\{
\begin{array}{ll}
y    & \mbox{if $x = \mbox{comp}_{M}(y)$} \\
\bot & \mbox{otherwise,} 
\end{array}
\right.
\]
where $\bot$ is a special symbol indicating,
in the usage ``$f(x) = \bot$,'' that $f$ on $x$ is not
defined.
Clearly, given~$x$, it can be checked in
polynomial time whether $x$ encodes an accepting path of $M$ (by
checking whether it starts with the initial configuration of $M$ for
some input string, all transitions from one configuration to the next
are legal, and the final configuration contains an accepting final
state), and if so, the input string $y$ of $M$ can easily be
determined.  Thus, $f \in \fp$.  Since $M$ is a UP machine, $f$ is
injective. The polynomial bounding the running time of $M$ witnesses
the honesty of~$f$. Since $L(M) = \sigmastar$, $f$ is surjective.
Finally, $f^{-1} \not\in \fp$, since $f^{-1}(y) = x$ is an accepting
computation of $M(y)$ for each~$y$, and so $f^{-1} \in \fp$
contradicts our assumption that $M$ only has hard certificates.  To
summarize, $f$ is a partial one-one one-way function with
$\mbox{range}(f) = \sigmastar$.

\smallskip

(2) implies (3): Let $f$ be a partial one-one one-way function with
$\mbox{range}(f) = \sigmastar$.  We will show that
\mbox{$\mbox{range}(f) = \sigmastar$} is not in $\easyforall (\up)$.
Let $p$ be the polynomial that witnesses the honesty of~$f$.  Consider
the following machine~$M$. On input~$y$, $M$ nondeterministically
guesses all strings $x$ of length at most~$p(|y|)$, computes $f(x)$
for each guessed~$x$, and accepts $y$ if and only if \mbox{$f(x) =
  y$}.  Clearly, $M$ is a UP machine accepting $\sigmastar$, since $f$
is a $p$-honest 
bijection (from some subset of $\sigmastar$ onto~$\sigmastar$)
computable in polynomial time.  Since $f^{-1} \not\in \fp$ and the
accepting path of $M(y)$ contains \mbox{$x = f^{-1}(y)$}, no FP
function can output, for all~$y$, the accepting path of $M$ on
input~$y$. Thus, $\sigmastar \not\in \easyforall (\up)$.~\hfill$\Box$

\medskip

By Grollmann and Selman's characterization of the existence of
partial one-one one-way functions with $\mbox{range}(f) =
\sigmastar$~\cite{gro-sel:j:complexity-measures}, we immediately have
Corollary~\ref{cor1:up}, which has 
previously been proven directly 
by Hartmanis and Hemaspaandra (then 
Hemachandra)~\cite{har-hem:j:up},
using different notation.
As a point of interest, we note that Corollary~\ref{cor1:up}
proves that separating P from a certain class containing P 
is equivalent to
separating P from a certain class contained in P\@. 
Also, though Naor and 
Impagliazzo~\cite[Proposition~4.2]{nao-imp:c:decision-trees}
(see 
also~\cite{cre-sil:c:sperner,for-rog:c:separability,fen-for-nai-rog:c:inverse})
have shown that for the converse of the original (i.e., NP) version of the 
Borodin-Demers~\cite{bor-dem:t:selfreducibility} 
theorem\footnote{\protect\singlespacing
Which 
says
$\p \neq \np \cap \conp$ implies $\easyforall \neq \p$,
except it states this in a different but equivalent form.}
there is a relativized counterexample,
Corollary~\ref{cor1:up} says that the converse of 
the UP analog of the Borodin-Demers theorem holds
(see~\cite{har-hem:j:up} for discussion of this point).

\begin{corollary} {} \cite{har-hem:j:up} \quad
\label{cor1:up}
$\p \neq \up \cap \coup \ $ if and only if $\ \easyforall (\up) \neq \p$.
\end{corollary}

A seemingly
unrelated connection comes from finite model theory.
Gr\"adel~\cite{gra:j:one-way} has recently shown that $\p = \up \cap
\coup \ $ if and only if the weak definability principle holds for every
first order logic ${\cal L}$ on finite structures that captures~P\@.
The weak definability principle says: Every totally defined query (on
the set of finite structures of the relations of a first order logic
${\cal L}$) that is implicitly definable in ${\cal L}$ is also
explicitly definable in ${\cal L}$ 
(see~\cite{gra:j:one-way} for
those notions not defined here).

\begin{corollary}
$\easyforall (\up) \neq \p\ $ if and only if the weak
  definability principle fails for some first order logic ${\cal L}$ on
  finite structures that captures P.
\end{corollary}

Fenner et al.~\cite{fen-for-nai-rog:c:inverse} also consider the
``one-bit version'' of the condition $\sigmastar \in \easyforall$. Let
us define $1\mbox{-}\easyforall ({\cal C})$ to be the class of all
sets $L$ that either are finite, or that 
satisfy (a)~$L \in {\cal C}$, and~(b) for
every ${\cal C}$-machine $N$ such that $L(N) = L$, there exists an FP
function $f_N$ such that, for all $x \in L$, $f_{N}(x)$ outputs the
first bit of an (``the'' in the 
case ${\cal C} = \up$) 
accepting path of~$N(x)$. Clearly (as in the case of
NP), we have for the UP case:
(a)~$\p = \easyforall (\up)$ implies $\p = 1\mbox{-}\easyforall (\up)$,
and
(b)~$\p = 1\mbox{-}\easyforall (\up)$
implies
$\p = \up \cap \coup$.
Thus, Corollary~\ref{cor1:up} in fact can be restated as
Corollary~\ref{cor2:up}, which sharply contrasts with the NP 
case~\cite{nao-imp:c:decision-trees,for-rog:c:separability,fen-for-nai-rog:c:inverse},
i.e., even though $\p = \easyforall$, $\p = 1\mbox{-}\easyforall$, and
$\p = \np \cap \conp$ appear to be pairwise different conditions,
their UP variants behave equivalently, and thus it is not reasonable to
consider a ``one-bit version'' of $\easyforall (\up)$.

\begin{corollary}
{} (see also \cite{har-hem:j:up}) \quad
\label{cor2:up}
The collapses $\p = \easyforall (\up)$, $\p = 1\mbox{-}\easyforall
(\up)$, and $\p = \up \cap \coup$ are pairwise equivalent.
\end{corollary}

Now we characterize the existence of surjective poly-one one-way
functions by separating P and $\easyforall (\fewp)$. 

\typeout{Note to us: Try to find 2-class separation characterizing these conditions}

\begin{theorem}
\label{thm:interm}
The following are equivalent.
\begin{enumerate}

\item There exists a partial surjective poly-one one-way function.

\item There exists a total surjective poly-one one-way function.

\item There exists a total poly-one one-way function $f$ with
  $\mbox{range}(f) \in \p$.

\item There exists a partial poly-one one-way function $f$ with
  $\mbox{range}(f) \in \p$.

\item $\easyforall (\fewp) \neq \p$.

\item $\sigmastar \not\in \easyforall (\fewp)$.

\item $\easyforall (\fewp)$ is not closed under complementation.
\end{enumerate}
\end{theorem}

\noindent {\bf Proof.} \quad 
Clearly, (1) implies~(3), as if $f$ is 
a function satisfying~(1), then 
\begin{eqnarray*}
g(x) & \equalsdef  & \left\{
\begin{array}{ll}
   0f(x) & \mbox{if
      $f(x) \neq \bot$} \\ 
    1x & \mbox{if
      $f(x) = \bot$}
\end{array}
\right.
\end{eqnarray*}
satisfies~(3). Also, (3) trivially implies~(4).

\smallskip

(4) implies (5): Let $f$ be a partial poly-one one-way function with
$\mbox{range}(f)$ in P\@.  We will show that \mbox{$\mbox{range}(f)$}
is not in $\easyforall (\fewp)$.  Let $p$ be the polynomial that
witnesses the honesty of~$f$.  Consider the following machine~$M$. On
input~$y$, $M$ nondeterministically guesses all strings $x$ of length
at most~$p(|y|)$, computes $f(x)$ for each guessed~$x$, and accepts
$y$ if and only if \mbox{$f(x) = y$}.  Clearly, $M$ is a FewP machine
accepting $\mbox{range}(f)$, since $f$ is a $p$-honest poly-one function
computable in polynomial time.  Since $f$ is not FP-invertible and
each accepting path of $M(y)$ contains some value of 
$f^{-1}(y)$, no FP function can output, for all~$y$, some accepting
path of $M$ on input~$y$. Thus, $\mbox{range}(f) \not\in \easyforall
(\fewp)$.

\smallskip

It is clear that~(2) implies~(1).  Suppose~(1) holds, and $f$ is a
function satisfying~(1).  Then $f'$ is a function satisfying~(2),
where 
\begin{eqnarray*}
f'(x) & \equalsdef  & \left\{
\begin{array}{ll}
   \epsilon & \mbox{if
      $x = \epsilon$} \\ 
    f(z)0 & \mbox{if
      $x=z0$ and $f(z) \neq \bot$} \\
    z1 & \mbox{if
      $x=z0$ and $f(z) = \bot$} \\
    z1 & \mbox{if
      $x=z1$.}
\end{array}
\right.
\end{eqnarray*}

\smallskip

The proof that conditions~(5),~(6), and~(7) of this theorem are pairwise
equivalent goes through as in the proof of the corresponding claim for
$\easyforall$ or $\easyforall (\up)$ (see Theorem~\ref{thm:up}).
Finally, that (7) implies (1) can again
be seen as in the proof of Theorem~\ref{thm:up}, the only difference
being that $M$ now is a FewP machine accepting $\sigmastar$ and the
function $f$ is now defined by \mbox{$f(x) = y$} if $x$ is {\em
  some\/} accepting path of $M(y)$, and $f(x)$ is undefined otherwise.
Then, $f$ is a partial surjective poly-one one-way function. This
completes the proof that all statements of the theorem are
equivalent.~\hfill$\Box$

\medskip

Note that $\p \neq \fewp$ is clearly implied by each of the 
conditions of Theorem~\ref{thm:interm}.  Note also that
$\p\neq\fewp \cap \cofewp$ clearly implies each
of the conditions of Theorem~\ref{thm:interm}, though it is not 
known whether the converse holds.  We conjecture that it does
not (equivalently, we conjecture that the converse of the FewP analog of 
the Borodin-Demers theorem does not hold).  Thus, the conditions
of Theorem~\ref{thm:interm} are intermediate 
between the conditions $\p \neq \fewp \cap  \cofewp$
and $\p \neq \fewp$.\footnote{\protect\singlespacing
\protect\label{foo:interm}
Regarding the condition $\p \neq \fewp$,
Allender~\protect\cite{all:coutdatedExceptForPUNCstuff:complexity-sparse}
showed that the following conditions are all equivalent:
(a)~$\p \neq \fewp$,
(b)~there exists a total poly-one one-way function, and
(c)~there exists a total poly-one weak one-way function.
Weak one-way functions mean the following.
A poly-one function $f$ is {\em strongly
    $\fp$-invertible\/} if there is a function $g \in \fp$ such that
  for every $y \in \mbox{range}(f)$, $g(y)$ prints {\em all\/} elements
  of $f^{-1}(y)$.
A function $f$ is called a 
   {\em weak one-way function\/} if $f \in \fp$, $f$
is poly-one, $f$ is honest, and $f$ is not
   strongly FP-invertible. 

Similarly, it is not hard to see, e.g., from Allender's proof,
that also equivalent to (a), (b), and (c) are each of 
these conditions:  (d)~
there exists a total poly-one weak one-way function $f$ with
  $\mbox{range}(f) \in \p$,
and
(e)~there exists a partial poly-one one-way function.

We note that the following condition is also equivalent to each 
of (a)--(e):  (f)~there exists a total surjective 
poly-one weak one-way function.  This is true for the following
reasons.  Clearly (f) implies~(d).  Also, (e)~implies (f)~as if $h$
is a function satisfying~(e), then $h'$ satisfies~(f), where
\begin{eqnarray*}
h'(x) & \equalsdef  & \left\{
\begin{array}{ll}
   \epsilon & \mbox{if
      $x \in \{\epsilon,0,1\}$} \\ 
    h(z)0 & \mbox{if
      $x=z00$ and $h(z) \neq \bot$} \\
    z1 & \mbox{if
      $(x=z00 \wedge h(z) = \bot)$ or $x=z11$} \\
    z0 & \mbox{if
      $x=z01$ or $x=z10$.}
\end{array}
\right.
\end{eqnarray*}
}

Could it be the case that the conditions of 
Theorem~\ref{thm:interm} in fact either are 
equivalent to $\p \neq \fewp \cap  \cofewp$, or are 
equivalent to
$\p \neq \fewp$?~~Relativized counterexamples are known
for each of these cases. 
In particular, 
there is a relativized world,
constructed
by Fortnow and Rogers~\cite{for-rog:c:separability}, 
in which the conditions of 
Theorem~\ref{thm:interm} fail yet 
$\p \neq \fewp$ holds.
Also, Lance Fortnow~\cite{for:perscomm:971210-oracle} has informed us
that, using the techniques of 
Fortnow and Rogers~\cite{for-rog:c:separability}, 
one can build a relativized world in which 
in which 
$\p = \fewp \cap \cofewp$
yet 
the conditions of 
Theorem~\ref{thm:interm} hold.

\section{Characterizing the Existence of One-Way Permutations}

For many types of one-way functions, the existence question has been
characterized in the literature as equivalent to the separation of
suitable complexity classes.  Such a characterization for the
existence of one-way permutations, however, is still missing.  To
date, the result closest to this goal is the above-mentioned
characterization of the existence of a {\em partial, injective, and
surjective\/} one-way function $f$ by the condition $\p \neq \up \cap
\coup$~\cite{gro-sel:j:complexity-measures}.\footnote{\protect\singlespacing
Fenner et al.~\cite{fen-for-nai-rog:c:inverse}
make the following claim: If $\p = 1\mbox{-}\easyforall$, then there
exist no one-way permutations. However, since $\p =
1\mbox{-}\easyforall$ implies $\p = \up \cap \coup$, the following
also correct claim is stronger: If $\p = \up \cap \coup$, then there
exist no one-way permutations. The difficult part seems to be the
converse implication, and we conjecture that the converse does not
hold.} 
Since $f$ is not total, $f$ is not a permutation of
$\sigmastar$ (even though $f$ is a bijection mapping a subset of
$\sigmastar$ onto $\sigmastar$). Thus, $\p \neq \up \cap \coup$
potentially is a strictly weaker condition than the existence of a
one-way permutation. Of course, such a function $f$ can be made
total~\cite{gro-sel:j:complexity-measures}, but only at the cost of
loss of surjectivity (even though such a total one-way function
created from $f$ still has a range in~P). However, we will show below
that the existence of one-way permutations is equivalent to the
existence of total injective one-way functions whose range is
P-rankable.

\begin{definition}
\cite{gol-sip:j:compression}
\quad
  A set $A$ is said to be P-{\em rankable\/} if there exists a
  polynomial-time computable function~{\em rank\/} so that
$
(\forall x \in \sigmastar)\, [\mbox{\em rank\/}(x) =
\|A^{\leq_{\mbox{\protect\tiny lex}} x}\|]
$,
where $A^{\leq_{\mbox{\protect\tiny lex}} x}$ denotes the set of all
strings $w \in A$ with $w \leq_{\mbox{\protect\scriptsize lex}} x$.
\end{definition}

That is, a ranking function for $A$ tells us the number of strings in
$A$ up to a given string. 
To avoid confusion, we mention that the notion of P-rankability 
used here (and in~\cite{gol-sip:j:compression})
is also sometimes referred to 
as ``strong P-rankability'' (e.g., in~\cite{hem-rud:j:ranking}). 

\begin{theorem}
\label{thm:permutation}
One-way permutations exist if and only if there exist total one-one
one-way functions whose range is P-rankable.
\end{theorem}

\noindent {\bf Proof.} \quad 
The ``only if'' direction is immediate, since $\sigmastar$ is P-rankable.

\smallskip

For the converse, suppose there exists a total one-one one-way
function $f$ whose range is P-rankable. We will define a one-way
permutation~$h$.  Intuitively, the idea is to fill in the
holes in the range of~$f$, using its P-rankability. Let $T =
\mbox{range}(f)$ be P-rankable.  For each~$n$, let
$
\mbox{\em holes\/}(n) \equalsdef 2^n - \|T^{=n}\|.
$
Note that since $T$ is P-rankable, {\em holes\/} is in~$\fp$. Let us
introduce some useful notation. For each string~$x$, let $k(x)$ be the
lexicographical position of $x$ among the length $|x|$ strings; e.g.,
$k(000) = 1$ and $k(111) = 8$. For each string~$x$ and each $j \in \N$,
let $x-j$ denote the string that in lexicographical order comes $j$
places before~$x$. For each set $A$ and each $k \in \N$, let $A_{[k]}$ be
the $k$th string of $A$ in lexicographical order. Now define the
function $h$ by
\begin{eqnarray*}
h(x) & \equalsdef & \left\{
\begin{array}{ll}
  f(x - \sum_{i=0}^{|x|} \mbox{\em holes\/}(i)) & \mbox{if
      $k(x) > \mbox{\em holes\/}(|x|)$} \\
  \left( \overline{T} \cap \Sigma^{|x|}\right)_{[k(x)]} & \mbox{if
      $k(x) \leq \mbox{\em holes\/}(|x|)$.}
\end{array}
\right.
\end{eqnarray*}
Since $T$ is P-rankable and $f \in \fp$, we have $h \in \fp$. Clearly,
$h$ is honest and injective, $h$ is total, and $\mbox{range}(h) =
\sigmastar$. If one could invert $h$ in polynomial time, then $f$
would also be FP-invertible, as the P-rankability of $T$ allows one to find the
string in the range of $f$ that should be inverted with respect to
$h$, and after inverting we shift the inverse with respect to~$h$,
say~$z$, by $\sum_{i=0}^{|z|} \mbox{\em holes\/}(i)$ positions to obtain the
true inverse with respect to~$f$. Hence, $h$ is a one-way
permutation.~\hfill$\Box$

\medskip

\begin{table}[tp]
\small
\begin{tabular}{|l||c|c|}\hline
{\em Partial functions} & one-one          & poly-one           \\ \hline\hline
no restriction   & $\p \neq \up$~\cite{gro-sel:j:complexity-measures}       & 
$\p \neq \fewp$ (Footnote~\ref{foo:interm}) \\ \hline
surjective       & 
$\p \neq \easyforall (\up )$ (Thm.~\ref{thm:up})
                          & 
$\p \neq \easyforall (\fewp )$ (Thm.~\ref{thm:interm})  \\ \hline
range in P       & 
$\p \neq \easyforall (\up )$ 
(Thm.~\ref{thm:up} plus 
\protect\cite[Theorem~8]{gro-sel:j:complexity-measures})     
                          &   
$\p \neq \easyforall (\fewp )$ (Thm.~\ref{thm:interm})    \\ \hline
\end{tabular}
\caption{Characterizations of the existence of various types of one-way 
functions: the partial function case.\label{t:X}}
\end{table}

\begin{table}[tp]
\small
\begin{tabular}{|l||c|c|}\hline
{\em Total functions}      & one-one              & poly-one \\ \hline\hline
no restriction   & 
$\p \neq \up$~\cite{gro-sel:j:complexity-measures}        
& $\p \neq \fewp$~\cite{all:coutdatedExceptForPUNCstuff:complexity-sparse}        \\ \hline
surjective       & 
{\rm open question (but note Thm.~\ref{thm:permutation})}  
& 
$\p \neq \easyforall (\fewp )$ (Thm.~\ref{thm:interm})
\\ \hline
range in P       & 
$\p \neq \easyforall (\up )$ 
(Thm.~\ref{thm:up} plus 
\protect\cite[Theorem~8]{gro-sel:j:complexity-measures})     
& 
$\p \neq \easyforall (\fewp )$ (Thm.~\ref{thm:interm}) \\ \hline
weak             &  
$\p \neq \up$~\cite{gro-sel:j:complexity-measures}        
   & 
$\p \neq \fewp$~\cite{all:coutdatedExceptForPUNCstuff:complexity-sparse}    \\ \hline
surj.\ \& weak   &  
{\rm open question (but note Thm.~\ref{thm:permutation})}  
& 
$\p \neq \fewp$ (Footnote~\ref{foo:interm}) \\ \hline
P-range \& weak  &  
$\p \neq \easyforall (\up )$ 
(Thm.~\ref{thm:up} plus 
\protect\cite[Theorem~8]{gro-sel:j:complexity-measures})     
& 
$\p \neq \fewp$ (Footnote~\ref{foo:interm}) \\ \hline
\end{tabular}
\caption{Characterizations of the existence of various types of one-way 
functions: the total function case.\label{t:Y}}
\end{table}

Note that P-rankability of the range of $f$ suffices to give us
Theorem~\ref{thm:permutation}, and Theorem~\ref{thm:permutation}
is stated in this way.
However, even weaker notions would
work.  Without going into precise details, we remark that one
just needs a function that, from some easily found and countable set
of places, is an honest {\em address function\/}
(see~\cite{gol-hem-kun:j:address}) for the complement of the range
of~$f$.
Of course, the ultimate goal is to find a characterization of the
existence of one-way permutations in terms of a separation of suitable 
complexity classes.

Finally, Tables~\ref{t:X} and \ref{t:Y}
summarize the characterization results that are known from the
literature and from this paper. Note that for one-one functions, 
FP-invertibility and strong
FP-invertibility are clearly identical notions, and so the one-one 
column of 
Tables~\ref{t:X} and~\ref{t:Y} is not affected by 
the ``weak'' issue.

{\samepage
\begin{center}
{\bf Acknowledgments}
\end{center}
\nopagebreak
\noindent We gratefully acknowledge interesting discussions with 
Erich Gr\"adel
and
Gerd Wechsung on this subject.  We are indebted 
to Lance Fortnow 
for the oracle discussion in the last 
paragraph of Section~\ref{s:sur-one},
and to Alan Selman for
generously permitting us to include his result that $\easyforall = \p$
is equivalent to the closure of $\easyforall$ under complement, which
led to Part~\ref{selman-item} of Theorem~\ref{thm:up}.
}%

{\singlespacing

{\bibliography{gry}}
}

\end{document}